\documentclass[twoside,final]{pro12}
\usepackage[english]{babel}
\usepackage[utf8]{inputenc}
\usepackage{amsmath}
\usepackage{graphicx}
\usepackage{array}
\usepackage{SIunits}
\usepackage{lscape}
\usepackage{multirow}
\usepackage{float}

\usepackage{lineno}

\newenvironment{correctiontwo}{}

\newenvironment{correctionthree}{}

\head{C.\,K. Louis et al.} {Jupiter--satellite DAM emissions: a parametric study}	

\begin{document}

\title{Simulating Jupiter--satellite decametric emissions with ExPRES: a parametric study}	
\author{C. K. Louis\adress{\textsl LESIA, Observatoire de Paris, PSL, CNRS, SU/UPMC, UPD, 5 place Jules Janssen, 92195 Meudon, France}$^*$\adress{\textsl USN, Observatoire de Paris, CNRS, PSL, UO/OSUC, Nan\c cay, France}$^\dag$, L. Lamy$^{*\dag}$, P. Zarka$^{*\dag}$, B. Cecconi$^{*\dag}$, \\ S.\,L.\,G. Hess\adress{\textsl ONERA / DESP, Toulouse, France}$\,$, and X. Bonnin$^*$}

\maketitle

\begin{abstract}
The high latitude radio emissions produced by the \correctiontwo{Cyclotron Maser Instability (CMI) in Jupiter's magnetosphere} extend from a few kHz to 40~MHz. Part of the decametric emissions is of auroral origin, and part is driven by the moons Io, Europa and Ganymede. After summarizing the method used to identify Jupiter--satellite radio emissions, which consists in comparing space-- and ground--based radio observations to ExPRES simulations of CMI--driven emissions in the time--frequency plane, we present a parametric study of the free parameters required by the ExPRES code (electron distribution function and resonant energy, magnetic field model, lead angle, \correctionthree{and} altitude of the ionospheric cut--off) in order to assess the accuracy of our simulations in the Io--Jupiter case. \correctiontwo{We find that Io--DAM arcs are \correctionthree{fairly} modeled by loss--cone driven CMI with electrons of 1--10 keV energy, using the ISaAC, VIPAL \correctionthree{or VIP4} magnetic field model and a simple sinusoidal lead angle model. The altitude of the ionospheric cut--off has a marginal impact on the simulations. We discuss the impact of our results on the identification of Europa--DAM and Ganymede--DAM emissions.}
\end{abstract}

\section{Introduction}

\correctiontwo{Jupiter's high latitude magnetospheric radio emissions are known to be} produced via the Cyclotron Maser Instability (CMI), in which non--Maxwellian weakly relativistic electrons gyrating along high latitude magnetic field lines amplify radio waves at a frequency close to the local electron cyclotron frequency [Zarka, 1998, Treumann, 2006, and references therein; Louarn et al., 2017]. These emissions are circularly or elliptically polarized extraordinary mode radiation, beamed along thin hollow conical sheet at large angle from the magnetic field line in the source. The Jovian radio spectrum is composed of several spectral components, including the kilometric (b--KOM), hectometric (HOM), and decametric (DAM) radiations. \correctiontwo{The latter is partly controlled by Io (Io--DAM) [Bigg, 1964], Europa (Europa--DAM) or Ganymede (Ganymede--DAM) [Louis et al., 2017a; Zarka et al., 2017, this issue], or \correctionthree{satellite--independent} [Kurth et al., 2017a]}.

\correctiontwo{DAM emission, typically observed between 1 and 40 MHz, is mainly structured in the form of discrete arcs in dynamic spectra, i.e. displays of the intensity distribution in the time--frequency (t--f) plane. These arcs were historically divided in categories labelled as A, B, C and D, depending on their occurrence in the (Io phase--Observer's longitude) plane [Carr et al., 1983]. These four categories were later identified to originate from two physical sources (A \& B, right--hand polarized, in the northern hemisphere, and C \& D, left--handed polarized, in the southern hemisphere). The emission of each source, anisotropically beamed in widely open conical sheets, is observed from the eastern or western limb of the planet [Hess et al., 2014, and references therein]. Emissions from the eastern limb (A \& C) appear as vertex--late arcs (i.e. closing parentheses) in dynamic spectra, whereas western limb (B \& D) emissions appear as vertex--early arcs (i.e. opening parentheses) [Marques et al., 2017, and references therein].}

\correctiontwo{To identify satellite--induced DAM emissions, Louis et al. [2017a,b] used the ExPRES code [Hess et al., 2008] to simulate their expected signature in the t--f plane, that they directly compared to radio dynamic spectra. The principle of the method (ExPRES code, inputs parameters, identification criteria) is summarized in Section \ref{method}. Section \ref{parametric} presents a parametric study of the input parameters based on typical Io--DAM arcs. It enables us to quantify the influence of each input parameter on the modeled arcs and to confirm the choice of parameters employed in recent studies, that provide simulations in fair agreement with the observations.}

\section{Detection of Jupiter--satellite induced radio emissions}
\label{method}

The method developed to search for DAM emissions induced by satellite--Jupiter interactions is fully described in [Louis et al., 2017a]. Hereafter we briefly outline the method and the identification criteria.

\subsection{Principle of ExPRES}

The Exoplanetary and Planetary Radio Emissions Simulator (ExPRES) \correctiontwo{(described in section 3 of Hess et al. [2008])} computes the geometrical visibility of \correctiontwo{sources of CMI--driven radio emission in the environment of a magnetized body}. Simulated radio sources are placed along selected magnetic field lines (e.g. the Io--Jupiter or nearby field lines), \correctionthree{in which} a source at frequency $f$ is located where $f \simeq f_{ce}$ \correctiontwo{(with $f_{ce}=\frac{eB}{2 \pi m}$ the electron cyclotron frequency, $B$ the local magnetic field amplitude, and $e$ and $m$ the electron charge and mass)} and where the condition $f_{pe}/f_{ce}<0.1$ \correctiontwo{(with $f_{pe} = \frac{1}{2 \pi} \sqrt{\frac{ne^2}{m \epsilon_0}}$ the local electron plasma frequency, $n$ the electron density, and $\epsilon_0$ the permittivity of free space)}, required for CMI--driven emission, is fulfilled [Hilgers, 1992; Zarka et al., 2001]. \correctiontwo{The ExPRES code computes the beaming angle $\theta$ at each frequency, and determines at each time step at which frequency -- if any -- the emission direction matches that of a given observer, fixed or moving, within a cone thickness $\delta \theta$ (generally assumed to be 1$\degree$).}

This produces t--f spectrograms of radio emission occurrence for each simulated source, that can be directly compared to observations. ExPRES was developed and used to simulate radio sources from Jupiter [Hess et al., 2008, 2010; Cecconi et al., 2012; Louis et al., 2017a,b], Saturn [Lamy et al., 2008, 2013], and exoplanets [Hess and Zarka, 2011].

The simulations of Io--DAM [Louis et al., 2017b], Europa--DAM and Ganymede--DAM [Louis et al., 2017b] were built as follows. We used the Jovian magnetic field model ISaAC (In--Situ and Auroral Constrains) [Hess et al., 2017, this issue], an updated version of the VIPAL model of Hess et al. [2011] further constrained by the locus of the UV auroral footprints of Europa and Ganymede. \correctiontwo{We added the contribution from the simple current sheet model of Connerney et al. [1981]. The magnetospheric plasma density is the sum of two contributions, from the planetary ionospheric [Hinson et al., 1998] and the Io plasma torus [Bagenal et al., 1994]. Simulated sources are placed over the frequency range 1--40 MHz along a source field line mapping the equatorial position of the Galilean satellite considered.}

\correctiontwo{The increased plasma density in the Io torus results in decreased Alfv\'en speed, thus Alfv\'en waves produced at Io need several tens of minutes to exit the torus [Saur et al., 2004], and eventually accelerate the electrons responsible for the radio emissions. This implies a longitudinal shift -- known as the lead angle $\delta$ -- between the magnetic field line connected to the moon and the radio--emitting field line. To simulate Io--DAM, we use the simple sinusoidal lead angle model of Hess et al. [2010]: }

\begin{equation}
\delta=A_{N/S}+B_{N/S} cos(\lambda_{Io}-202\degree)
\label{equationLAG}
\end{equation}

\correctiontwo{with $\lambda_{Io}$ the \correctionthree{westward} jovicentric longitude of Io, $A_N=2.8 \degree$ and $B_N=-3.5 \degree$ in the northern hemisphere (resp. $A_S=4.3 \degree$ and $B_S=3.5 \degree$ in the southern hemisphere).}

The emission (or beaming) angle $\theta$ was derived as in Hess et al. [2008] by assuming oblique CMI emission from a loss cone electron distribution function (found to better fit the observed arcs than the perpendicular emission from a shell distribution), with resonant electron energies $E_{e^-} \simeq 1$ keV in the northern hemisphere and $3$ keV in the southern one. 

\correctionthree{$\theta$ depends on the electron energy, likely to vary over time and space \correctiontwo{[Hess et al., 2010]}, and on the position of the sources, i.e. on the satellite considered, the magnetic field model and the lead angle $\delta$. Thus we needed to define criteria to match simulated and observed arcs.}

\subsection{Identification of Jupiter--satellite emissions}
\label{criteria}
\correctiontwo{In Louis et al. [2017a,b] we performed extensive simulations of Jupiter--satellite emissions using the above--described input parameters, and compared them to ground and space--based observations. In the first study, we used Voyager and Cassini observations to identify Europa--DAM and Ganymede--DAM emissions. In the second one, we identified Io--DAM arcs in simultaneous Juno, Nan\c{c}ay and Wind observations, with shapes very close to the simulated ones.}

\correctiontwo{To account for the necessarily imperfect fit between modeled and observed arcs, we only retained the candidate arcs fulfilling the following criteria:
 the observed emission had to be (1) a well identified single structure (i.e. clearly distinguishable among surrounding emissions), (2) an arc with the same curvature as the simulated one (vertex--late or vertex--early), (3) close enough in time (within a window of $\pm 2$ hours for Io, $+2$/$-5$ hours for Europa, and $+2$/$-8$ hours for Ganymede, see next paragraph), (4) continuously extending over a bandwidth $\geq$3 MHz and with a maximum frequency $\geq$5 MHz, and (5) with a polarization consistent with the predicted one. Finally (6) candidate arcs were discarded if they re--occurred $\sim9$ h $55$ m earlier or later, as they were then attributed to auroral radio sources co--rotating with Jupiter.}

The above time windows result from two causes of uncertainty: (i) the beaming angle $\theta(f)$ which, \correctiontwo{in the frame of loss cone--driven CMI, depends} on the electron energy, the magnetic field model (that constrains the source position) and \correctionthree{the peak auroral altitude $h_{\text{max}}$ (taken as the mean value of the mirror point, below which precipitating electrons are lost by collision)} (see Eq. 4 of Hess et al. [2008]), and (ii) the uncertainty on the lead angle $\delta$.

\section{Simulations of Io--induced radio emissions: a parametric study}
\label{parametric}
Here we focus on the effect of the different \correctiontwo{input parameters of ExPRES on the modeled dynamic spectra. These parameters are the electron distribution function, the magnetic field model, the model of the lead angle $\delta$ and, in the loss cone case, the energy of resonant electrons and \correctionthree{the peak auroral altitude} $h_{\text{max}}$.} We study 4 typical Io arcs, one of each category (A,B,C,D), observed by Juno/Waves (Figure \ref{Io-C-D}) [Kurth et al., 2017b] and Voyager/PRA (Figure \ref{Io-A-B}) [Warwick et al., 1977] over their full spectral range.

\subsection{Io--DAM from Jupiter's southern hemisphere}

\begin{figure}[H]
\centering
\includegraphics[width=0.9\textwidth]{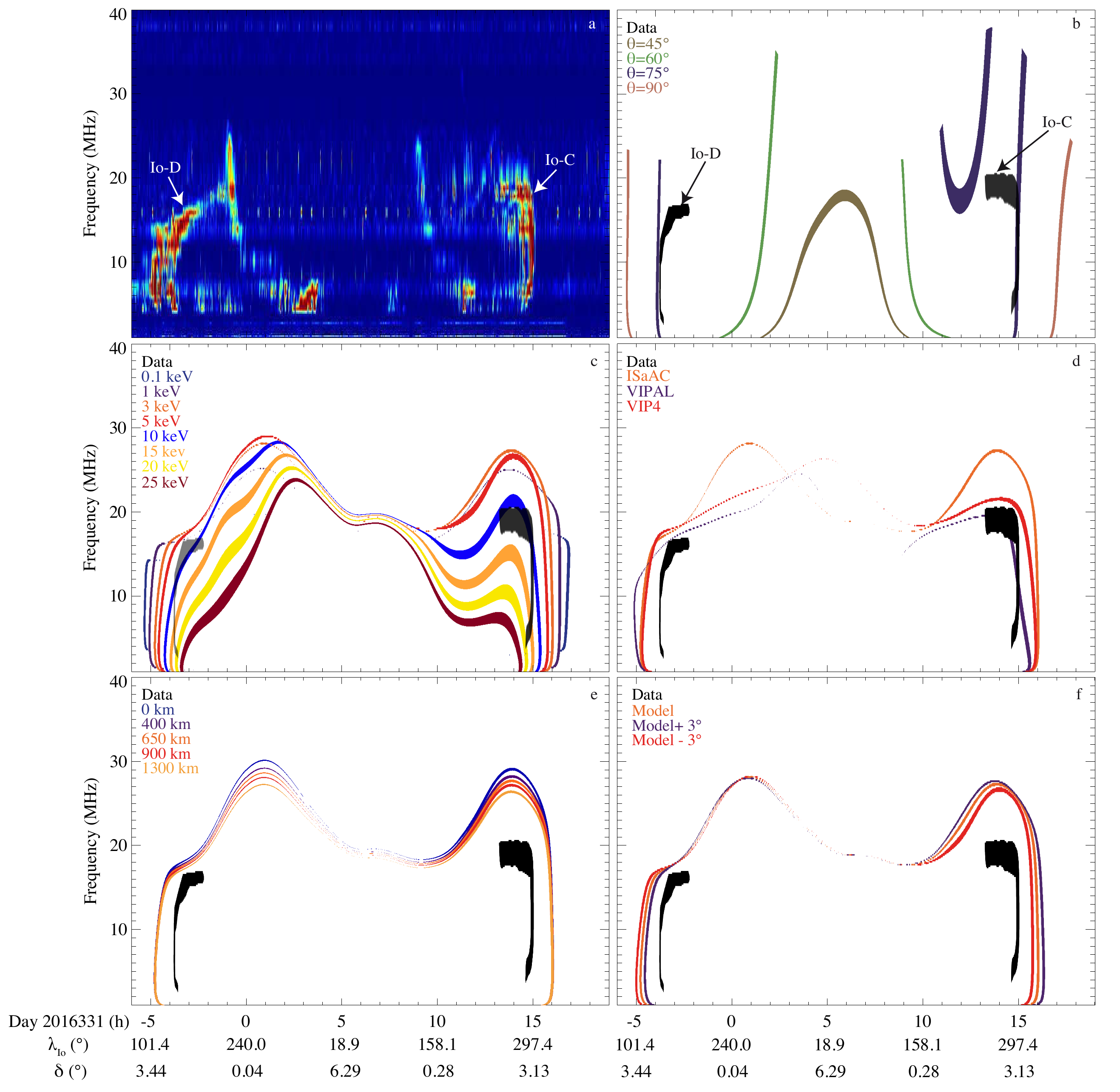}
\caption{(a) Juno/Waves dynamic spectrum of Io--C and Io--D arcs, whose masks are reproduced in black in panels (b--f) aside ExPRES simulations. 
In panels (b--f), one parameter is varied relative to the reference simulation described in the text.
(b) The 4 colored areas display simulated arcs obtained with a beaming angle constant in frequency, ranging from $45 \degree$ to $90 \degree$. Although the simulation with $\theta = 75 \degree$ matches part of the observed arcs, none of them correctly reproduces their overall shape.
(c) The 8 colored areas display simulated arcs obtained with electron energies between $0.1$ and $25$ keV. The blue simulation ($E_{e^-} = 10$ keV) best fits both Io--D and --C arcs above 10 MHz.
(d) The 3 colored areas correspond to different magnetic field models. Arcs modeled with VIP4 and ISaAC best fit the observed Io--D arc, while VIPAL provides a best fit to the observed Io--C arc.
(e) The 5 colored areas explore the value of $h_{\text{max}}$ between $0$ and $1300$ km above the 1 bar level. All simulations are very similar, a value of $h_{\text{max}}=900$ to $1300$ km making them slightly closer to the observed arcs.
(f) The orange simulation, obtained with a lead angle $\delta$ given by the model of Eq. \ref{equationLAG}, is surrounded by simulations obtained with modified models $\delta+ 3\degree$ (violet) and $\delta-3\degree$ (red). The modified model reduce the time shift for one arc and increase it for the other, bringing no significant improvement.
In addition to the time (in hours of Day of Year 2016331), Io's longitude $\lambda_{\text{Io}}$ and the corresponding lead angle $\delta$ are indicated in abscissa.}
\label{Io-C-D}
\end{figure}

Figure \ref{Io-C-D} shows \correctiontwo{6 dynamic spectra: observed Io--C and Io--D arcs (a) are compared to 5 sets of simulations, each exploring the variations of a single ExPRES input parameter (b--f). The reference simulation, from which parameters are varied, is based on loss cone--driven CMI with 3 keV electrons, the ISaAC magnetic field model, the lead angle model from (Eq. \ref{equationLAG}), and $h_{\text{max}} = 650$ km.}

\correctiontwo{An occurrence map (or mask) of each arc was extracted from panel (a) by manually selecting the contour of the t--f pixels from each arc with an intensity larger than the median intensity level at each frequency. These masks are reproduced in black on panels (b--f).}

\correctiontwo{Table \ref{table1} lists the time delays $\Delta T$ between the mask of each observed and each simulated arc. $\Delta T_\text{Mean}$ is calculated as the arithmetic mean of the time delays between the observed and simulated arcs at each frequency, between $f_\text{min}=4$ MHz and the lowest maximum frequency of the two arcs, $f_\text{max}$. $\Delta T_\text{MAD}$ is mean absolute deviation between the observed and simulated arcs, computed as the mean of the absolute time delays between the observed and simulated arcs at each frequency, in the same range as above:}

\begin{equation}
\Delta T_\text{MAD}=\frac{1}{n_{\text{freq}}} \sum_{i=4 \text{~MHz}}^{f_\text{max}} |t_{\text{obs}}(i)-t_{\text{sim}}(i)|
\label{tabs}
\end{equation}

These two estimates $\Delta T$ provide complementary information. \correctiontwo{$\Delta T_\text{Mean}$ characterizes the temporal accuracy of the simulation: if $\Delta T_\text{Mean}<0$ the simulation arc occurs too early, whereas if $\Delta T_\text{Mean} > 0$ it occurs too late. Used together, both $\Delta T$ values characterize the fidelity of the t--f shape of the simulated arc: if both $\Delta T \simeq 0$ then the simulation correctly fits the observed arc along their common frequency range; if $\Delta T_\text{Mean} \simeq 0$ but $\Delta T_\text{MAD} \neq 0$ then the shape is not well simulated although the occurrence time is correct. Both $\Delta T$ are listed in Table 1 for all simulations of Figure \ref{Io-C-D}.}

\begin{table}[]
\footnotesize
\centering
\begin{tabular}{c|c|cc|cc|cc|cc}
		&			& \multicolumn{2}{c|}{Io--A (1979048)}	& \multicolumn{2}{c|}{Io--B (1979209)}	& \multicolumn{2}{c|}{Io--C (2016331)}	& \multicolumn{2}{c}{Io--D (2016330)} \\ 
		&			& \multicolumn{2}{c|}{$\Delta T$ (hours)}	& \multicolumn{2}{c|}{$\Delta T$ (hours)}	& \multicolumn{2}{c|}{$\Delta T$ (hours)}	& \multicolumn{2}{c}{$\Delta T$ (hours)} \\ 
		& 			& $MAD$ & $Mean$ 				& $MAD$ & $Mean$ 				& $MAD$ & $Mean$ 				& $MAD$ & $Mean$ \\\hline
		& $45 \degree$	& 3.63	& -3.63               & 9.36             & 9.36                & 7.22             & -7.22               & 7.10             & 7.10                \\
Constant	& $60 \degree$	& 3.63	& -3.63               & 2.51             & 2.51                & 5.58             & -5.58               & 4.88             & 4.88                \\
$\theta$	& $75 \degree$	& 0.17	& 0.17                & 1.07             & 1.07                & 0.19             & 0.13                & 0.28             & -0.28               \\
		& $90 \degree$	& 1.36	& 1.36                & 0.86             & -0.86               & 3.04             & 3.04                & 1.83             & -1.83               \\\hline
		& 0.1 		& 1.51	& 1.51                & 0.40             & -0.40               & 1.86             & 1.85                & 1.57             & -1.57               \\
		& 1 keV		& 1.10	& 1.10                & 0.09             & -0.09               & 1.26             & 1.26                & 0.94             & -0.94               \\
		& 3 keV		& 0.73	& 0.73                & 0.25             & 0.25                & 0.84             & 0.84                & 0.52             & -0.52               \\
		& 5 keV		& 0.51	& 0.51                & 0.45             & 0.45                & 0.63             & 0.63                & 0.35             & -0.35               \\
$E_{e^-}$	& 10 keV		& 0.27	& 0.05                & 0.92             & 0.92                & 0.21             & 0.18                & 0.18             & 0.00                \\
		& 15 keV		& 0.29	& -0.26               & 1.19             & 1.19                & 0.47             & -0.41               & 0.49             & 0.48                \\
		& 20 keV		& 0.53	& -0.53               & 1.34             & 1.34                & 2.17             & -2.17               & 1.43             & 1.43                \\
		& 25 keV		& 0.76	& -0.76               & 1.44             & 1.44                & 2.88             & -2.88               & 2.20             & 2.20                \\\hline
		& ISaAC		& 1.10	& 1.10                & 0.09             & -0.09               & 0.84             & 0.84                & 0.52             & -0.52               \\
B model	& VIP4		& 1.20	& 1.20                & 0.06             & 0.06                & 0.79             & 0.79                & 0.60             & -0.60               \\
		& VIPAL		& 0.82	& 0.82                & 0.49             & -0.49               & 0.24             & 0.12                & 0.87             & -0.87               \\\hline
		& 0 km		& 1.12	& 1.11                & 0.09             & -0.09               & 0.84             & 0.84                & 0.52             & -0.52               \\
		& 400 km		& 1.11	& 1.11                & 0.09             & -0.09               & 0.84             & 0.84                & 0.52             & -0.52               \\
$h_{\text{max}}$ & 650 km & 1.10	& 1.10                & 0.09             & -0.09               & 0.84             & 0.84                & 0.52             & -0.52               \\
		& 900 km		& 1.10	& 1.09                & 0.09             & -0.08               & 0.84             & 0.84                & 0.51             & -0.51               \\
		& 1300 km	& 1.09	& 1.09                & 0.08             & -0.08               & 0.83             & 0.83                & 0.51             & -0.51               \\\hline
		& Model		& 1.10	& 1.10                & 0.09             & -0.09               & 0.84             & 0.84                & 0.52             & -0.52               \\
$\delta$	& Model $+ 3 \degree$ & 1.29	& 1.28                & 0.15             & 0.15                & 1.01             & 1.01                & 0.40             & -0.40               \\
		& Model $- 3\degree$  & 0.92	& 0.92                & 0.29             & -0.29               & 0.67             & 0.67                & 0.63             & -0.63               \\ 
\end{tabular}
\caption{Time shifts in hours between emission contours and simulations for all test parameters of Figures \ref{Io-C-D} and \ref{Io-A-B}. $\Delta T_{MAD}$ is the mean absolute deviation, and $\Delta T_{mean}$ the arithmetic mean, both computed from the time delay measured at each frequency between the observed and simulated arcs (from $f_\text{min}=4$ MHz to the lowest maximum frequency of the two arcs, $f_\text{max}$).}
\label{table1}
\end{table}

\correctiontwo{Figure \ref{Io-C-D}b tests four different lead angles $\theta$, kept constant as a function of frequency, including $\theta=90\degree$ that corresponds to shell--driven CMI, and oblique emission at $\theta=45\degree$, $60\degree$ and $75\degree$. No simulation correctly fits the observed arcs, but for $\theta=75\degree$ the simulations match the Io--D arc in the $4-12$ MHz range and the Io--C arc in the $6-19$ MHz range. This shows that Io--DAM emission has an oblique beaming, with $\theta(f)$, reaching $\simeq 75\degree$ at $\simeq 5-10$ MHz, and decreasing towards lower and higher frequencies, as predicted by loss cone--driven CMI [Hess et al., 2008; Ray and Hess, 2008].}

\correctiontwo{Figure \ref{Io-C-D}c therefore tests the loss cone electron distribution function by varying the resonant electron energy from $E_{e^-} = 0.1$ to $25$ keV. The simulations with $1$ to $15$ keV energy better fit the global shape of both arcs ($\Delta T_\text{MAD} \simeq |\Delta T_\text{Mean}|$), albeit with a modest temporal accuracy ($-0.94$ h $\leq \Delta T_\text{Mean} \leq 1.26$ h). Overall, the best fit of both arcs, for the observation studied, is obtained with $E_{e^-} = 10$ keV ($\Delta T_\text{Mean} = 0.00$ and $\Delta T_\text{MAD} = 0.18$). A lower value such as $3$ keV, slightly modifies the time shift ($\Delta T_\text{MAD} = 0.84$ h for Io--C and $0.52$ h for Io--D emission), that remains well within our search window of $\pm 2$ h, but it also induces a larger mismatch in maximum frequency ($\sim$6 MHz for the Io--C arc).}

\correctiontwo{Figure \ref{Io-C-D}d tests the three magnetic field models VIP4, VIPAL and ISaAC, the former two having been previously used in Io--DAM simulations [Hess et al., 2010, 2011]. The simulated arcs are separated by less than $0.6$ h in $\Delta T_\text{MAD}$, but they display larger differences in maximum frequency (up to $7$ MHz for Io--D). For the observation studied, the best fit is provided by VIPAL for Io--C ($\Delta T_\text{MAD}=0.24$ h, $\Delta T_\text{Mean}=0.12$ h), and by ISaAC and VIP4 for Io--D ($|\Delta T| = 0.52$ h and $0.60$ h respectively). So no model provides a strong advantage for the simulations.}

Figure \ref{Io-C-D}e \correctiontwo{tests the $h_{\text{max}}$ parameter, ranging from $0$ km to $1300$ km [Hinson et al., 1998, Bonfond et al., 2009]. It only slightly changes the peak frequency of the simulations (by $<3$ MHz). Computed time shifts remain within $0.01$ h of each other for all simulations. Thus the $h_{\text{max}}$ parameter does not play a significant role. The altitude of the ionospheric cut--off has thus a marginal impact on the simulations, a value about 900--1300 km leading to a very slightly better match.}

Finally, figure \ref{Io-C-D}f \correctiontwo{varies the lead angle (indicated on the x--axis), testing the sinusoidal model of Hess et al. [2010] (Eq. \ref{equationLAG}), and this model shifted by $\pm 3\degree$ (i.e. $\simeq 50$\% of its maximum value). Time shifts are slightly modified in the latter cases (by less than $\pm$0.2 h, i.e. well within our search window of $\pm 2$ h) but in opposite ways for Io--C and Io--D arcs, thus bringing no net improvement to the unmodified model. The difference on the peak frequency of the simulations is also very small ($\leq 1$ MHz).}

\subsection{Io--DAM from Jupiter's northern hemisphere}

\begin{figure}[H]
\centering
\includegraphics[width=0.9\textwidth]{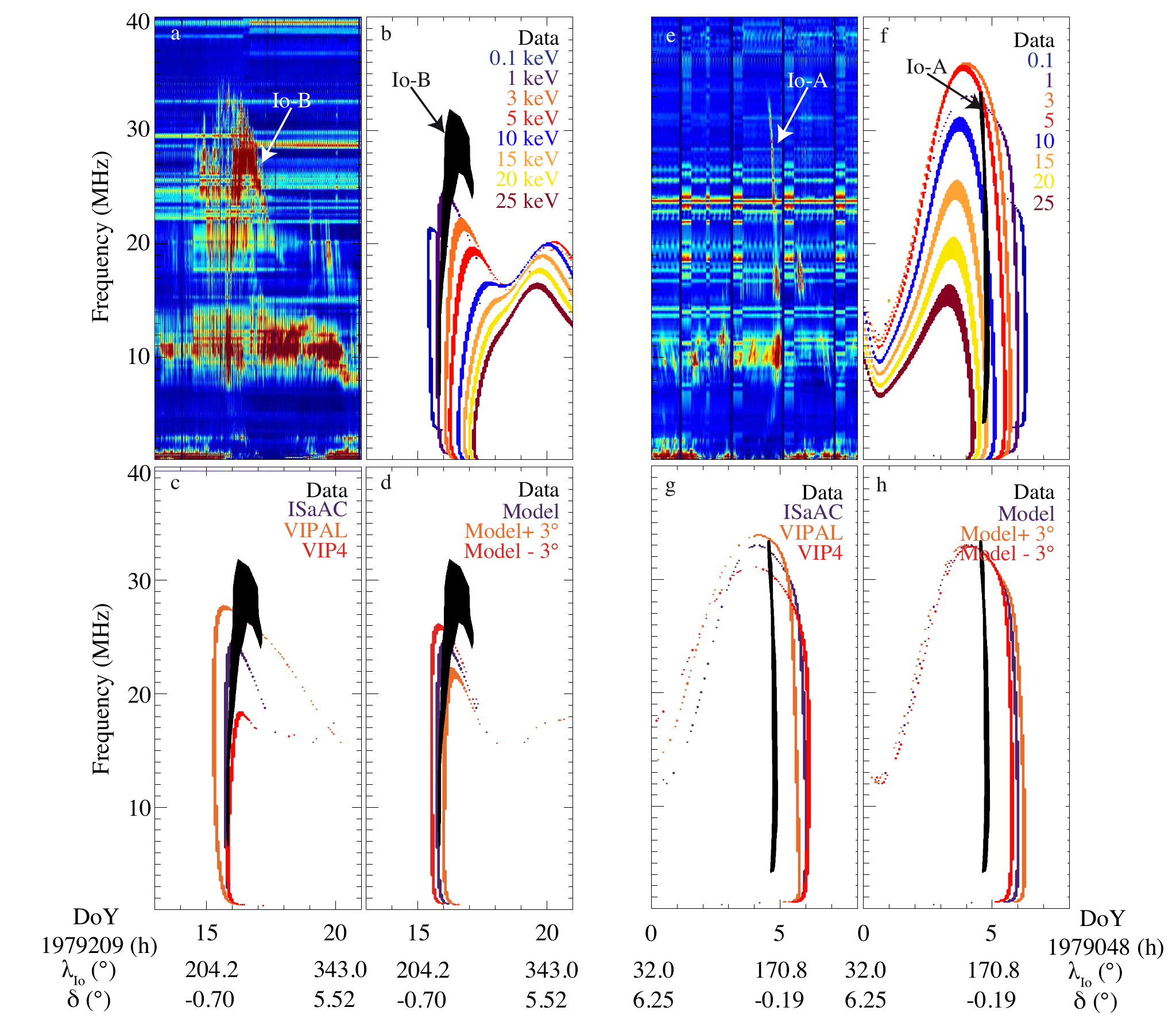}
\caption{(a) Voyager 2/PRA observation of an Io--B arc, whose mask is reproduced in black in panels (b--d) aside ExPRES simulations. (e) Voyager 1/PRA observation of an Io--A arc, and (f--h) corresponding ExPRES simulations. The reference simulation is the same as for Figure \ref{Io-C-D}, but with $1$ keV electrons.
(b \& f) The 8 colored areas display simulated arcs obtained with electron energies between 0.1 and 25 keV. The colored areas display simulations using electrons energies ranging $0.1$ to $25$ keV. The violet simulation ($E_{e^-} = 1$ keV) best fits the observed Io--B arc up to 25 MHz, while the best fit for the Io--A arc is obtained with $E_{e^-} =$ 1--10 keV.
(c \& g) The 3 colored areas correspond to different magnetic field models. The Io--B arc modeled with ISaAC best fits the observed one. For the Io--A arc, the three simulations occur too late, VIPAL doing slightly better than the other models.
(d \& h) The simulations test the lead angle model $\delta$ of Eq. \ref{equationLAG} (violet), and the modified models $\delta+ 3\degree$ (orange) and $\delta-3\degree$ (red). For the Io--B arc, the unmodified model best fits the emission, while for the Io--A arc, all three models poorly match the observed emission.
In addition to the time (in hours of DoY 1979209 for the Io--B, and DoY 1979048 for the Io--A arc), Io's longitude $\lambda_{\text{Io}}$ and the corresponding lead angle $\delta$ are indicated in abscissa.}
\label{Io-A-B}
\end{figure}

\correctiontwo{In this section, we investigate northern Io--DAM emissions observed by Voyager/PRA. Figure \ref{Io-A-B}a,e display typical examples of Io--B and Io--A arcs. Their masks, reproduced in black in all other panels, were extracted as in Figure \ref{Io-C-D} but with a lower intensity threshold in order to cope with the lower SNR, interference and variable spectral response resulting from the antenna resonance (we used the 5\% quantile at 289 kHz). Panels (b--d) and (f--h) test the influence of the same parameters as Figure \ref{Io-C-D}c,d,f, which have the strongest impact on simulated arcs. The reference simulation, from which parameters are varied, is the same as in section 3.1, but with $1$ keV electrons. The computed time delays $\Delta T_\text{MAD}$ and $\Delta T_{mean}$ are also listed in Table \ref{table1}.}

\correctiontwo{Figures \ref{Io-A-B}b,f test the loss cone electron distribution function by varying the resonant electron energy from $E_{e^-}=0.1$ to $25$ keV. The simulations with 1 to 10 keV energy better fit the shape of the observed Io--A arc while those with 1 keV energy (and up to 3 keV) better fit the observed Io--B arc ($\Delta T_\text{MAD} \simeq |\Delta T_\text{Mean}|$). However the temporal accuracy of the simulations is modest, as 0.05 h $\leq \Delta T_\text{Mean} \leq$ 1.10 h for Io--A and -0.40 h $\leq \Delta T_\text{Mean} \leq$ 0.45 h for Io--B. Overall, the best fits for these observations are obtained with $E_{e^-}=10 $ keV for the Io--A arc ($\Delta T_\text{MAD}=0.27$ h and a $\Delta T_\text{Mean}=0.05$ h), and with $E_{e^-}=1 $ keV for the Io--B arc ($\Delta T_\text{MAD}= |\Delta T_\text{Mean}|=0.09$ h). A lower value in the Io--A case, such as 1 keV, increases the time shifts to 1.10 h, remaining within our search window of $\pm 2$ h, but it also induces a larger mismatch in maximum frequency (up to 5 MHz).}

\correctiontwo{Figures \ref{Io-A-B}c,g test the three magnetic field models. The simulated arcs are separated by less than 0.4 h in $\Delta T_\text{MAD}$, but they display larger differences in maximum frequency (up to $9$ MHz for Io--B between VIP4 and VIPAL). For the observations studied, the best fits are provided by VIPAL for Io--A ($\Delta T_\text{MAD}=\Delta T_\text{Mean}=0.82$ h), and by ISaAC for Io--B ($\Delta T_\text{MAD}= |\Delta T_\text{Mean}|=0.09$ h). For Io--B, the time shifts are slightly smaller with VIP4, but the error on the maximum frequency is larger. ISaAC and VIPAL thus do a better job, although \correctionthree{it is} still far from perfect.}

\correctiontwo{Finally, Figures \ref{Io-A-B}d,h test the lead angles. The simulated arcs are again separated by less than 0.4 h for Io--A and 0.2 h for Io--B, well within our search window. For the Io--B arc, the unmodified model fits slightly better the emission, while for the Io--A arc, all three models poorly match the observed emission. The difference on the peak frequency of the simulations is also small ($\leq 4$ MHz for Io--B).}

\section{Discussion}

\subsection{Standard input parameters for modeling Io--DAM}

\correctiontwo{Our parametric study clearly demonstrates that an oblique and variable beaming angle $\theta(f)$ is required to simulate correctly Io--DAM emissions, fully consistent with loss cone--driven CMI. The altitude of the ionospheric cut--off $h_\text{max}$ has a marginal impact on the simulations, with differences in time shifts $\leq 0.03$ h. None of the three magnetic field models clearly stands out as better than the others, as they lead to differences in time shifts $\leq 0.6$ h. The numerical values of Table \ref{table1} are not sufficient to \correctionthree{quantify the accuracy of the fit of the arc shape}, as the time shifts are computed over the common frequency ranges of the observed and simulated arcs (above 4 MHz). Frequency coverage is thus an important parameter. ISaAC and VIPAL reach higher frequencies than VIP4, in better agreement with the observations in the northern hemisphere. In the southern hemisphere, ISaAC predicts too high maximum arc frequencies, so that the preference goes to VIPAL and VIP4. Juno/MAG observations indeed revealed a magnetic field close to the planet (radius $<1.3$ $R_J$, i.e. where most of the decameter sources above ~10 MHz lie) dramatically different from that predicted by existing spherical harmonic models [Connerney et al., 2017]. We have to wait for a better magnetic field model.}

\correctiontwo{The most important parameters appear to be the lead angle $\delta$ and the energy of the resonant electrons $E_{e^-}$. Albeit not perfect (in part due to the imperfect magnetic field models), the sinusoidal variation of $\delta$ described in Eq. \ref{equationLAG} is probably a good approximation of the real variation of $\delta$, and a global shift by $\pm3\degree$, that induces differences in time shifts $\leq 0.5$ h, does not bring any significant improvement to the quality of the fits. About electron energy, we find that the best results are obtained for $E_{e^-}$ in the range 1--15 keV in the south and 1--10 keV in the north. For the best fit values of the studied cases ($E_{e^-}=10$ keV for Io--A, --C and --D arcs, and $E_{e^-}=1$ keV for the Io--B arc), the associated time shifts are within $-0.09$ h $\leq \Delta T_\text{Mean} \leq 0.18$ h.}

\correctiontwo{The electron energy and the plasma density in the Io torus (and hence the lead angle) are likely to vary over time. This makes difficult to choose \textit{a priori} the best parameters for simulating a specific arc. In order to perform statistical studies of many arcs, we thus need to define a set of reference parameters and hypotheses for running extensive simulations. In Louis et al. [2017b] we chose (1) the loss cone--driven CMI, with (2) the ISaAC model, (3) the model lead angle of Hess et al. [2010], (4) $h_\text{max}=650$ km, and (5) a resonant electrons energy of $1$ keV in the northern hemisphere and $3$ keV in the southern one. Our parametric study suggests that these energies are reasonable but on the low side, but they led us to a good fit of all Io--DAM arcs observed by Juno/Waves, well within a search window of $\pm 2$ h.}

\subsection{Application to Europa--DAM and Ganymede--DAM}

\correctiontwo{Louis et al. [2017a] applied ExPRES simulations to the tentative identification of Europa--DAM and Ganymede--DAM emissions in Voyager/PRA and Cassini/RPWS observations. The simulations are based on the same parameters and hypotheses as above, except for the lead angle $\delta$ that deserved further thinking, as there is no structure comparable to Io's torus along the orbits of Europa and Ganymede.}

\correctiontwo{For Ganymede, Bonfond et al. [2013] noticed multiple spots of the Ganymede northern UV auroral footprint with a maximum longitudinal shift of $13 \degree$. This translates into a maximum lead angle $\delta_{Ga}=13 \degree$, corresponding to a maximum temporal delay of $-6$h15m. Bonfond et al. [2013, 2017] concluded that the multiple Ganymede spots could results from an increase of the Jovian plasma sheet at Ganymede's orbit. But as these multiple UV spots are not always observed, the plasma sheet must be very variable, and so will be the lead angle $\delta_{Ga}$, that cannot be simulated in a deterministic way. In the absence of a similar study of Europa multiple UV footprints, we simply assumed a maximum lead angle $\delta_{Eu} = 10\degree$, intermediate between those of Io and Ganymede, that corresponds to a maximum delay of $-2$h20m. We assumed for $\delta_{Eu}$ and $\delta_{Ga}$ a variation of the same form as Eq. \ref{equationLAG}, replacing $\lambda_{Io}$ by $\lambda_{Eu}$ or $\lambda_{Ga}$, with parameters $A$ and $B$ chosen for having $0\degree \leq \delta_{Eu} \leq 10\degree$ and $0\degree \leq \delta_{Ga} \leq 13\degree$. When $\delta_{Eu},\delta_{Ga}=0\degree$, the temporal search window for matching ExPRES simulations and observations is thus $\pm 2$h as for Io--DAM (due to the uncertainty on the electron energy and the other input parameters), and it extends to $+2$h$/-4$h20m for Europa when $\delta_{Eu}=10\degree$ and $+2$h$/-8$h15m for Ganymede when $\delta_{Ga}=13\degree$. These longer windows allow us to take into account possible increases of the plasma sheet density.}

\correctiontwo{These parameters and search windows allowed Louis et al. [2017a] to identify about a hundred Europa--DAM arcs, and as many Ganymede--DAM arcs. Each identification is not 100\% certain, but their statistics revealed a clear organization\correctionthree{, not only as a function of the orbital phase of the moons (partially induced by our selection criteria), but also as a function of the observer's longitude and the longitude of these moons}, which strongly supports the statistical significance of these detections. \correctionthree{Indeed, the Io--DAM emissions are organized as a function of the Io's longitude [Marques et al., 2017], as the Ganymede--DAM emissions, detected by Zarka et al. [2017], are organized as a function of the Ganymede's longitude. As a comparison, the auroral, satellite--independent, DAM emissions have an homogeneous distribution as a function of the moons' longitude, and are organized as a function of the observer's longitude only [Marques et al., 2017].}}

\section{Conclusions and Perspectives}
\label{conclusions}

With this parametric study we now have an overview of the \correctiontwo{influence of ExPRES inputs parameters on Jupiter--satellite simulated arcs. Our results can be summarized as follows: (1) $\theta(f)$ needs to be oblique and to decrease with frequency, which is achieved by assuming loss cone--driven CMI with resonant electron energies of a few keV, as previously shown by Hess et al. [2008] and Ray and Hess [2008].} (2) The altitude of the ionospheric cut--off plays a marginal role, only modifying the maximum arc frequency by $\leq 3$ MHz. (3) \correctiontwo{The three magnetic field models (VIP4, VIPAL, ISaAC) are roughly equivalent for Io--DAM simulations, leading to differences in time shifts $\leq 0.6$ h between the most separated simulated arcs, and differences in maximum arc frequency up to 9 MHz. However, by construction, the ISaAC model is better adapted to the Europa--DAM and Ganymede--DAM simulations.}

The most important free parameters are the electrons energy and the lead angle model. The electron energy responsible to the DAM emission is \correctiontwo{still poorly constrained} and is likely to vary over time, as the density of the Io torus. Thus, in the absence of \textit{a priori} information on the electrons energy in the radio sources, and on the plasma density along the moon orbit, it is reasonable for extensive simulation studies as in Louis et al. [2017a,b] to choose a constant electron energy and a model of lead angle, provided that we use a temporal search window of $\pm 2$ h for matching simulated and observed Io--DAM emissions. For Europa and Ganymede, the search window must be enlarge in correspondence with the expected lead angle variation, that depends on these moons' longitudes and on the variable plasma sheet along their orbit. The maximum search window is $+2$h$/-4$h20m for Europa and $+2$h$/-8$h15m for Ganymede.

Juno should enable the construction of a much more accurate magnetic field model, especially close to the planet in decameter radio sources, \correctionthree{and then better constrain their positions. The energy of the resonant electrons will be the main free parameter for matching simulations and observations. The potentially crossing by Juno of the moons' flux tubes (or tail) could inform us on the electron population, and will allow us to better constrain the fidelity of the simulated arc shapes.}

\textit{Acknowledgements:} The authors thank the anonymous referees for their thorough reviews, the Juno mission team --especially the Juno Waves instrument team--, and the MIRIADE service of IMCCE (especially J. Berthier) for the planet, satellites and spacecraft ephemeris. The ExPRES tool and the Kronos database are part of the MASER (Measurement, Analysis and Simulations of Emissions in Radio) service, supported by PADC (Paris Astronomical Data Center). Rehabilitation of the Voyager/PRA data was supported by CNES (Centre National d'\'Etudes Spatiales), that also supports the authors for their work on Juno. C. Louis is funded by the LABEX Plas@par project, managed by the Agence Nationale de la Recherche as a programme "Investissement d'avenir" with reference ANR--11--IDEX--0004--02.

\section*{References}
\everypar={\hangindent=1truecm \hangafter=1}
Bagenal,~F., Empirical model of the Io plasma torus: Voyager measurements, \textsl{J. Geophys. Res.}, \textbf{99}, 11043--11062, 1994.

Bigg,~E.\,K., Influence of the satellite Io on Jupiter's decametric emission, \textsl{Nature}, \textbf{203}, 1008--1010, 1964.

Bonfond,~B., D.~Grodent, J.--C.~G\'erard, A.~Radioti, V.~Dols, P.\,A.~Delamere, and J.\,T.~Clarke, The Io UV footprint: Location, inter-spot distances and tail vertical extent, \textsl{J. Geophys. Res.}, \textbf{114}, A07224, 2009.

Bonfond,~B., S.~Hess, J.--C.~G\'erard, D.~Grodent, A.~Radioti, V.~Chantry, J.~Saur, S.~Jacobsen, and J.\,T.~Clarke, Evolution of the Io footprint brightness I: Far--UV observations, \textsl{Planet. Space Sci.}, \textbf{88}, 64--75, 2013.

Bonfond,~B., D.~Grodent, S.\,V.~Badman, J.~Saur, J.--C.~G\'erard, A.~Radioti, Similarity of the Jovian satellite footprints: Spots multiplicity and dynamics, \textsl{Icarus}, \textbf{292}, 208--2017, 2017.

Carr,~T.\,D., M.\,D.~Desch, and J.\,K.~Alexander, Phenomenology of magnetospheric radio emissions, in \textsl{Physics of the Jovian Magnetosphere}, edited by A.\,J.~Dessler, Cambridge University Press, New York, USA, 226--284, 1983.

Cecconi,~B., et al. (10 co-authors), S.~Hess, A.~H\'erique, M.\,R.~Santovito, D.~Santos--Costa, P.~Zarka, G.~Alberti, D.~Blankenship, J.--L.~Bougeret, L.~Bruzzone, and W.~Kofman, Natural radio emission of Jupiter as interferences for radar investigations of the icy satellites of Jupiter, \textsl{Planet. Space Sci.}, \textbf{61}, 32--45, 2012.

Connerney,~J.\,E.\,P., The magnetic field of Jupiter, A generalized inverse approach, \textsl{J. Geophys. Res.}, \textbf{86}, 7679--7693, 1981.

Connerney,~J.\,E.\,P., et al. (21 co-authors), Jupiter's magnetosphere and aurorae observed by the Juno spacecraft during its first polar orbits, \textsl{Science}, \textbf{356}, 6340, 826--832, 2017.

Hess,~S.\,L.\,G. and P.~Zarka, Modeling the radio signature of the orbital parameters, rotation, and magnetic field of exoplanets, \textsl{Astro. Astrophys.}, \textbf{531}, A29, 2011.

Hess,~S.\,L.\,G., B.~Cecconi, and P.~Zarka, Modeling of Io--Jupiter decameter arcs, emission beaming and energy source, \textsl{Geophys. Res. Lett.}, \textbf{35}, L13107, 2008.

Hess,~S.\,L.\,G., A.~P\'{e}tin, P.~Zarka, B.~Bonford, and B.~Cecconi, Lead angles and emitting electron energies of Io--controlled decameter radio arcs, \textsl{Planet. Space Sci.}, \textbf{58}, 1188--1198, 2010.

Hess,~S.\,L.\,G., B.~Bonfond, P.~Zarka, and D.~Grodent, Model of Jovian magnetic field topology constrained by the Io auroral emissions, \textsl{J. Geophys. Res.}, \textbf{116}, A05217, 2011.

Hess,~S.\,L.\,G., E.~Echer, P.~Zarka, L.~Lamy, and P.\,A.~Delamere, Multi--instrument study of the Jovian radio emissions triggered by solar wind shocks and inferred magnetospheric subcorotation rates, \textsl{Planet. Space Sci.}, \textbf{99}, 136--148, 2014.

Hess,~S.\,L.\,G., B.~Bonfond, F.~Bagenal, and L.~Lamy, A model of the Jovian internal field derived from in--situ and auroral constrains, \textsl{Planetary Radio Emissions VIII}, edited by G. Fischer et al., Austrian Academy of Sciences Press, Vienna, this issue, 2017.

Hilgers,~A., The auroral radiating plasma cavities, \textsl{Geophys. Res. Lett.}, \textbf{19(3)}, 237--240, 1992.

Hinson,~D.\,P., J.\,D.~Twicken, and E.\,T.~Karayel, Jupiter's ionosphere: New results from Voyager 2 radio occultation measurements, \textsl{J. Geophys. Res.}, \textbf{103}, A5, 9505-9520,1998.

Kurth,~W.\,S., et al. (15 co-authors), M.~Imai, G.\,B.~Hospordarsky, D.\,A.~Gurnett, P.~Louarn, P.~Valek, F.~Allegrini, J.\,E.\,P.~Connerney, B.\,H.~Mauk, S.\,J.~Bolton, S.\,M.~Levin, A.~Adriani, F.~Bagenal, G.\,R.~Gladstone, D.\,J.~McComas, and P.~Zarka, A new view of Jupiter's auroral radio spectrum, \textsl{Geophys. Res. Lett.}, \textbf{44}, 7114--7121, 2017a.

Kurth,~W.\,S., G.\,B.~Hospodarsky, D.\,L.~Kirchner, B.\,T.~Mokrzycki, T.\,F.~Averkamp, W.\,T.~Robison, C.\,W.~Piker, M.~Sampl, and P.~Zarka, The Juno Waves Investigation, \textsl{Space Sci. Rev.}, 2017b.

Lamy,~L., P.~Zarka, B.~Cecconi, S.~Hess and R.~Prang\'e, Modeling of Saturn kilometric radiation arcs and equatorial shadow zone, \textsl{J. Geophys. Res.}, \textbf{113}, A10213, 2008.

Lamy,~L., R.~Prang\'e, W.~Pryor, J.~Gustin, S.\,V.~Badman, H.~Melin, T.~Stallard, D.\,G.~Mitchell, and P.\,C.~Brandt, Multispectral simultaneous diagnosis of Saturn's aurorae throughout a planetary rotation, \textsl{J. Geophys. Res.}, \textbf{118}, 4817--4843, 2013.

Louarn,~P., et al. (15 co-authors), F.~Allegrini, D.\,J.~McComas, P.\,W.~Valek, W.\,S.~Kurth, N.~Andr\'e, F.~Bagenal, S.~Bolton, J.~Connerney, R.\,W.~Ebert, M.~Imai, S.~Levin, J.\,R.~Szalay, S.~Weidner, R.\,J.~Wilson, and J.\,L.~Zink, Generation of the Jovian hectometric radiation: first lessons from Juno, \textsl{Geophys. Res. Lett.}, \textbf{44}, 4439--4446, 2017.

Louis,~C.\,K., L.~Lamy, P.~Zarka, B.~Cecconi, and S.\,L.\,G.~Hess, Detection of Jupiter decametric emissions controlled by Europa and Ganymede with Voyager/PRA and Cassini/RPWS, \textsl{J. Geophys. Res. Space Physics}, \textbf{122}, DOI: 10.1002/2016JA023779, 2017a.

Louis,~C.\,K., et al. (11 co-authors), L.~Lamy, P.~Zarka, B.~Cecconi, M.~Imai, W.\,S.~Kurth, G.~Hospodarsky, S.\,L.\,G.~Hess, X.~Bonnin, S.~Bolton, J.\,E.\,P.~Connerney, and S.\,M.~Levin, Io--Jupiter decametric arcs observed by Juno/Waves compared to ExPRES simulations, \textsl{Geophys. Res. Lett.}, \textbf{44}, DOI: 10.1002/2017GL073036, 2017b.

Marques,~M., P.~Zarka, E.~Echer, V.\,B.~Ryabov, M.\,V.~Alves, L.~Denis, and A.~Coffre, Statistical analysis of 26 years of observations of decametric radio emissions from Jupiter \textsl{Astron. Astrophys.}, \textbf{604}, A17, 2017.

Ray,~L.\,C. and S.\,L.\,G.~Hess, Modelling the Io--related DAM emission by modifying the beaming angle \textsl{J. Geophys. Res.}, \textbf{113}, A11218, 2008.

Saur,~J., B.\,H.~Mauk, A.~Ka{\ss}ner, and F.\,M.~Neubauer, A model for the azimuthal plasma velocity in Saturn's magnetosphere, \textsl{J. Geophys. Res.}, \textbf{109}, A05217, 2004.

Treumann, ~R.\,A., The electron--cyclotron maser for astrophysical application \textsl{Astron. Astrophys. Rev.}, \textbf{13}, 229--315, 2006.

Warwick,~J.\,W., J.\,B.~Pearce, R.\,G.~Peltzer, and A.\,C.~Riddle, Planetary Radio Astronomy experiment for Voyager missions, \textsl{Space Sci. Rev.}, \textbf{21}, 309--327, 1977.

Zarka,~P., Auroral radio emissions at the outer planets: Observations and theories, \textsl{J. Geophys. Res.}, \textbf{103}, 20159--20194, 1998.

Zarka,~P., J.~Queinnec, and F.\,J.~Crary, Low-frequency limit of Jovian radio emissions and implications on source locations and Io plasma wake, \textsl{Planet. Space Sci.}, \textbf{49}, 1137--1149, 2001.

Zarka,~P., M.\,S.~Marques, C.~Louis, V.\,B.~Ryabov, L.~Lamy, E.~Echer and B.~Cecconi, Radio emission from satellite-Jupiter interactions (especially Ganymede), \textsl{Planetary Radio Emissions VIII}, edited by G. Fischer et al., Austrian Academy of Sciences Press, Vienna, this issue, 2017.

\end{document}